\documentclass[prl,twocolumn,nofootinbib,aps,preprintnumbers,superscriptaddress]{revtex4-2}
\usepackage{graphicx}
\usepackage{color}
\usepackage{bm}
\usepackage{amsmath}
\usepackage{amssymb}
\usepackage{xspace}
\usepackage[normalem]{ulem}
\usepackage{units}
\usepackage{dcolumn}
\usepackage{paralist}
\usepackage{natmove}
\usepackage{natbib}
\usepackage{soul}

\usepackage{hyperref}

\begin{document}

\title{Bidirectional Learning of Relationships between Atomic Environments and \\ Electronic Band Dispersion in Semiconductor Heterostructures
}
\author{Artem~K.~Pimachev}
\author{Sanghamitra~Neogi}
\email{sanghamitra.neogi@colorado.edu}
\affiliation{Ann and H.J. Smead Aerospace Engineering Sciences, University of Colorado Boulder, Boulder, Colorado 80303, USA \\ 
Email: sanghamitra.neogi@colorado.edu} 




\begin{abstract}
Atomic-scale variations in semiconductor heterostructures, arising from strain, interfaces, and compositional modulation, strongly influence electronic band dispersion but remain difficult to probe and compare using first-principles methods alone. Here, we introduce a bidirectional learning approach that links local atomic environments to electronic band dispersion using atomically resolved spectral functions as information-dense representations. This formulation enables a forward model that predicts how atomic environments shape electronic bands, and a reverse model that infers atomic-environment descriptors directly from band dispersion images, including angle-resolved photoemission spectra. Applied to silicon/germanium superlattices and heterostructures, the approach reveals how inner and interfacial atomic environments give rise to distinct spectral signatures. The coupled forward–reverse framework enables self-consistent validation by reconstructing electronic band structures from inferred descriptors. By treating electronic bands as decomposable, learnable objects, this work provides a physics-informed route for interpreting spectroscopic data and for data-driven exploration of electronic properties in complex semiconductor heterostructures.

\end{abstract}

\maketitle

Semiconductor heterostructures, consisting of two or more layers of dissimilar semiconductor materials, are key platforms in condensed matter physics and electronic device applications~\cite{alferov2001nobel}. A primary advantage of these structures lies in their tunable physical properties, including electrical, magnetic, and optical responses that are governed by their electronic band structures. The electronic bands are, in turn, strongly influenced by the atomic arrangement within and across constituent layers. First-principles methods, particularly density functional theory (DFT), have been widely used to predict the electronic bands of complex materials. Accurate modeling of heterostructures typically requires large supercells to capture structural variations and interface effects, which incurs significant computational cost and limits systematic exploration of atomic configurations. Moreover, interpreting the electronic bands of heterostructures remains challenging: the bands may retain Bloch-like character of individual components or exhibit strong hybridization and mixing depending on atomic-scale structural features.

Although several band unfolding and spectral function techniques have been developed to analyze the character of electronic bands in disordered systems and alloys~\cite{ku2010unfolding,popescu2012extracting,boykin2007brillouin,boykin2009non,lee2013unfolding,chen2018layer,popescu2010effective,boykin2007approximate}, their application to heterostructures remains limited, and these approaches are fundamentally one-directional. Recent machine learning (ML) approaches have also demonstrated the ability to learn and reconstruct electronic band dispersion by mapping between different band representations~\cite{xian2023machine}. While such methods improve band interpretation and representation, they operate primarily at the level of global band structure and do not resolve contributions from local atomic environments. Crucially, no existing first-principles method provides a rapid and scalable means to predict band structures for arbitrary layered heterostructures or to invert the problem by inferring atomic-environment information directly from observed band dispersion. As a result, materials development for heterostructures continues to rely on costly trial-and-error cycles involving design, synthesis, and characterization. This gap motivates physics-informed learning approaches that establish direct links between atomic structure, electronic band dispersion, and functional properties, thereby enabling an inverse design paradigm.

In this work, we present an ML-assisted, first-principles-based approach that establishes a bidirectional relationship between atomic environments and electronic band dispersion in semiconductor heterostructures. Our approach combines two complementary learning models: (1) {\em a forward model} that predicts electronic band dispersion from atomic-environment descriptors, and (2) {\em a reverse model} that infers atomic-environment descriptors from band dispersion images. The reverse model is trained exclusively on DFT-computed spectral data, yet can process independent experimental angle-resolved photoemission spectroscopy (ARPES) images and predict plausible atomic environments. The approach preserves interpretability by leveraging atomically resolved spectral functions, which link local atomic environments to specific electronic dispersion features. Atomic descriptors inferred from band dispersion can guide the design of heterostructures with targeted electronic properties, enabling efficient exploration of structure–property relationships beyond conventional trial-and-error approaches. More broadly, because the spectral function representation operates directly on momentum-resolved electronic structure, it provides a physically grounded basis for scalable and transferable models of electronic band dispersion across diverse crystalline materials systems.

\section{Results}

\subsection{Forward model}

\subsubsection{Overview} 

\begin{figure}[h]
\begin{center}
\includegraphics[width=0.9\linewidth]{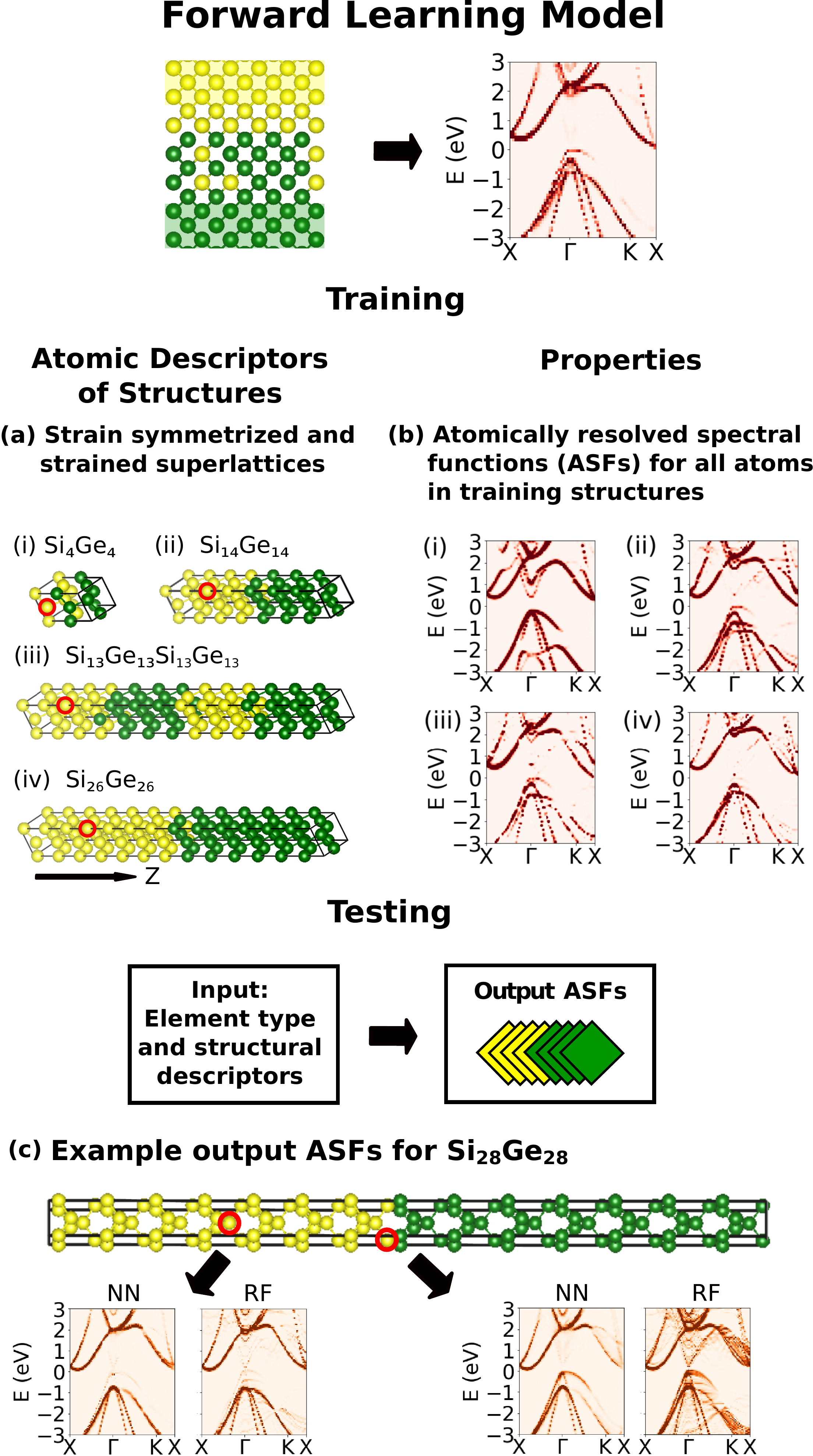}
\caption{{\bf Outline of the forward learning approach.} (a) Training data consist of strain-symmetrized and strained Si/Ge superlattices spanning a range of layer periods and compositions, including representative examples: (i) Si$_4$Ge$_4$, (ii) Si$_{14}$Ge$_{14}$, (iii) Si$_{13}$Ge$_{13}$Si$_{13}$Ge$_{13}$ and (iv) Si$_{26}$Ge$_{26}$. Atomic environments are described using element type and local structural descriptors. (b) Atomically resolved spectral functions (ASFs) for atoms highlighted in (a) (red circles), illustrating how distinct local atomic environments give rise to diverse electronic band dispersion features across superlattices. (c) Forward model predictions for a Si$_{28}$Ge$_{28}$ superlattice not included in the training set, demonstrating the model’s ability to generalize beyond the training data.
}
\label{fig:forward_ML}
\end{center}
\end{figure}

Figure~\ref{fig:forward_ML} outlines the forward learning approach used in this work to examine how local atomic environments influence electronic band dispersion in semiconductor heterostructures. The approach is built around a representation in which electronic structure is expressed using atomically resolved spectral functions (ASFs), enabling momentum-resolved band dispersion to be described within a common Brillouin-zone framework across superlattices and heterostructures of different periods and compositions. We demonstrate the approach using silicon/germanium (Si/Ge) superlattices and heterostructures (Fig.~\ref{fig:forward_ML}(a)), including both strain-symmetrized and strained configurations. In these models, superlattices consist of alternating Si and Ge layers, whereas heterostructures comprise multiple Si and Ge layers of varying thicknesses within a single supercell period. All supercells are periodically extended. Considering both strain-symmetrized and strained Si$_n$Ge$_n$ ($n$ denotes the number of monolayers) superlattices~\cite{d2012genetic,zhang2013genetic} (Table~\ref{table:MLDatatable}) allows us to capture the influence of epitaxial strain on electronic band dispersion~\cite{satpathy1988electronic, hybertsen1987theory, tserbak1993unified} and transport properties~\cite{proshchenko2019heat, proshchenko2019optimization,proshchenko2021role,settipalli2020theoretical,pimachev2021first,settipalli2022effect}. We describe atomic environments using elemental identity and local structural descriptors that encode coordination, bonding geometry, and directional information. Representative ASFs for selected atoms are shown in Fig.~\ref{fig:forward_ML}(b), illustrating systematic variations in band dispersion associated with different local atomic environments. These atom-resolved spectral signatures, together with the atomic-environment descriptors, define the input–output representation used to learn relationships between local structure and electronic band dispersion. The following subsections examine the physical insights revealed by ASFs and assess the predictive behavior and generalization of the forward model.

\begin{table*}[t]
\begin{center}
\caption{{\bf Summary of datasets used in the forward and reverse learning framework.}}
\centering 
\renewcommand{\arraystretch}{0.4}
\begin{tabular}{l l l l l}
\hline\hline \\                  
\textbf{Structure} & \textbf{Training} & \textbf{Features} & \textbf{Properties} & \textbf{Test}  \\ [1ex]
\textbf{Type} & \textbf{Structures} & & & \textbf{Structures}  \\ [1ex]
\hline \\                 
\multicolumn{5}{c}{Forward Learning Model: Neural Network (NN) \& Random Forests (RF) Model}\\ [1ex]
\hline  \\
Strain- & \ \ Si$_{2p}$Ge$_{2p}$ & $\bullet$ Atom type:  & $\bullet$  Spectral weights, & $\bullet$ HS: Si$_{8}$Ge$_{8}$Si$_{20}$Ge$_{20}$ \\ [1ex]
symmetrized &$ \ \ (p=1,2,\dots, 13)$  & \ \ 1 feature/atom  & \ \ $A^{p}(k,E)$: & \ \ 56 atoms  \\ [1ex]
and strained & \ \ (Si$_{2q-1}$Ge$_{2q-1}$)$^2$ & $\bullet$ Effective bond &  \ \ $ k \times E $ & \ \ Input features: \\ [1ex]
SLs &\ \ $(q=1,2,\dots, 7)$ & \ \ lengths, $b_{x}$ \& $b_{z}$: & \ \ $ = 64 \times 96 $ & \ \ 56 $\times$ 9 = 504 \\ [1ex]
& $\bullet$ 5 applied strains: & \ \ 2 features/atom & \ \ = 6144 & \ \ Output weights:\\ [1ex]
& \ \ [0.00\%, 0.59\%, & $\bullet$ Order parameters, & \ \ per atom ($p$) & \ \ 56 $\times$ 6144 = 344,064\\ [1ex]
& \ \ 1.16\%, 1.73\%, 2.31\%]  & \ \ $Q^{1,2,3}_{x, z}$: & $\bullet$ Total weights: & $\bullet$ SL: Si$_{28}$Ge$_{28}$ (SI Fig.~3) \\ [1ex]
& $\bullet$ Total: 120 structures &  \ \ 6 features/atom & \ \ $ 3360 \times 6144 $ &  \ \ 56 atoms \\ [1ex]
& $\bullet$ Number of atoms: & $\bullet$ Total: &  \ \ = 20,643,840 & \ \ Input features: 56 $\times$ 9 = 504  \\ [1ex]
& \ \ $6 \times 4 \times \Bigl(\sum_{i=1}^{13} p_i $& \ \ $3360 \times 9$ & &  \ \ Output weights:  \\[1ex]
& \ \ $ + \sum_{j=1}^7 (2q_j -1) \Bigl)$ & \ \ = 30,240 features & & \ \ 56 $\times$ 6144 = 344,064 \\ [2ex]
& \ \ = 3360 &  & & \ \ (Both strain-symmetrized)  \\ [1ex]
\hline  \\
\multicolumn{5}{c}{Reverse Learning Model: Convolutional Neural Network (CNN) Model}\\ [1ex]
\hline \\
Strain- & \ \ Same as Forward & $\bullet$ Spectral weights, & $\bullet$ Atom type: & $\bullet$ HS: Si$_{8}$Ge$_{8}$Si$_{20}$Ge$_{20}$ \\ [1ex]
symmetrized & \ \ Learning Model & \ \ $A_{E,k}$: & \ \ 1 feature/atom & \ \ Input ASFs pixels: \\ [1ex]
and strained & & \ \ $k \times E = 64 \times 64 $ & $\bullet$ Effective bond  & \ \ 56 $\times$ 64 $\times$ 64 \\ [1ex]
SLs &   & \ \ per atom & \ \ lengths, $b_{x}$ \& $b_{z}$:  & \ \ Output features: 56 $\times$ 9 \\ [1ex]
& & $\bullet$ Fermi level & \ \ 2 features/atom & $\bullet$ SL: Si$_{28}$Ge$_{28}$ (SI Fig.~5) \\ [1ex]
& &  \ \ alignments: & $\bullet$ Order parameters, & \ \ Input ASFs pixels: \\ [1 ex]
&  & \ \ 13 values around & \ \ $Q^{1,2,3}_{x, z}$: &  \ \ 56 $\times$ 64 $\times$ 64 \\ [1 ex]
& & \ \ -0.5 eV to +0.5 eV  & \ \ 6 features/atom  & \ \ Output features: 56 $\times$ 9 \\ [1 ex]
& & \ \  of mid-gap level & $\bullet$ Total: & $\bullet$ Bulk Si systems (SI Fig.~6-7) \\ [1 ex]
& & \ \ with step of 1/13 eV & \ \ $3360 \times 9$ &   $\bullet$ ARPES Si thin film \\ [1 ex]
& & $\bullet$ Total: & \ \ = 30,240 features & \ \ For all bulk cases  \\ [1ex]
& & \ \ $3360 \times 13 \times 64 \times 64$ &  & \ \ Input ASFs pixels: 64 $\times$ 64 \\ [1ex]
& & \ \ =$\underbrace{43,680}_{images}\times \underbrace{64 \times 64}_{pixels}$  &  & \ \ Output features: 9 \\ [1ex]
\hline \\
\multicolumn{5}{l}{(Si$_{2q-1}$Ge$_{2q-1}$)$^2$ $\equiv$ Si$_{2q-1}$Ge$_{2q-1}$Si$_{2q-1}$Ge$_{2q-1}$ for odd $q=1,2,\dots,7$}\\ [1ex]
\multicolumn{5}{l}{SL: Superlattice; HS: Heterostructure;} \\ [1ex]
\hline \\
\multicolumn{5}{c}{Combined Forward-Reverse Learning Framework: NN, RF \& CNN}\\ [1ex]
\hline \\
\multicolumn{5}{c}{$\bullet$ Relaxed and 1.73\% strained bulk Si} \\ [1 ex]
\multicolumn{5}{c}{CNN Model: Input pixels: 64 $\times$ 64; Output features: 9}\\ [1 ex]
\multicolumn{5}{c}{NN and RF Model: Input features: 9; Output weights: 64 $\times$ 96}\\ [1 ex]
\multicolumn{5}{c}{$\bullet$ Si ARPES spectra} \\ [1 ex]
\multicolumn{5}{c}{CNN Model: Input pixels: 64 $\times$ 64; Output features: 9}\\ [1 ex]
\multicolumn{5}{c}{NN and RF Model: Input features: 9; Output weights: 64 $\times$ 96}\\ [1 ex]
\hline \\
\end{tabular}
\label{table:MLDatatable}
\end{center}
\end{table*}

\subsubsection{Relationship between atomic-environment-descriptors and ASFs}

\begin{figure*}
\begin{center}
\includegraphics[width=1\linewidth]{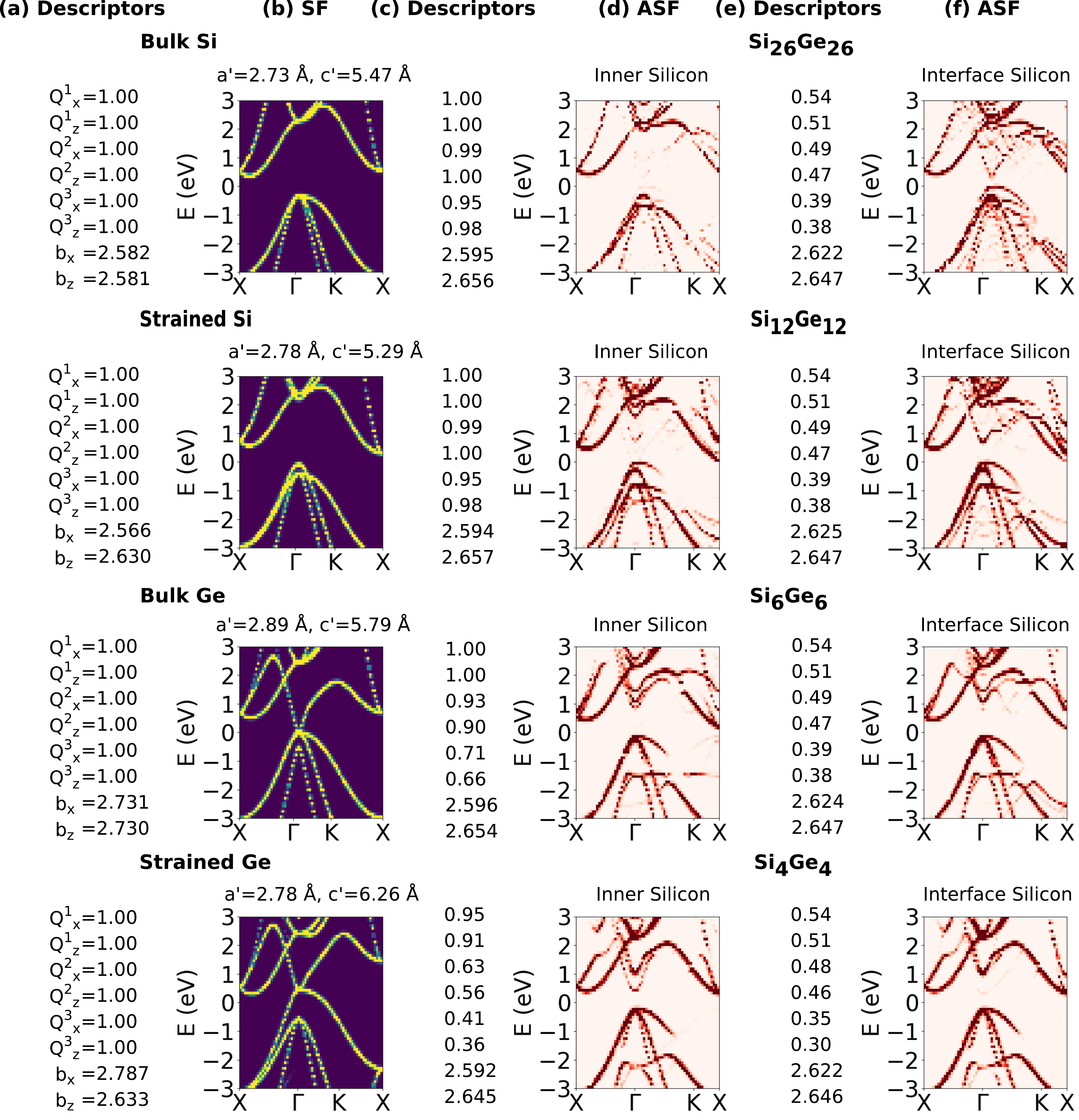}
\caption{{\bf Relationships between atomic-environment descriptors and spectral functions (SFs) in Si/Ge systems.} (a) Atomic-environment descriptors and (b) SFs for relaxed bulk Si (row 1), strained bulk Si (row 2), relaxed bulk Ge (row 3) and strained bulk Ge (row 4). (c,d) Descriptors and ASFs for inner Si atoms and (e,f) interface Si atoms in Si$_{26}$Ge$_{26}$ (row 1), Si$_{12}$Ge$_{12}$ (row 2), Si$_6$Ge$_6$ (row 3) and Si$_4$Ge$_4$ (row 4) superlattices.}
\label{fig:progression}
\end{center}
\end{figure*}

To establish the physical basis underlying the forward learning model, we examine how atomic-environment descriptors correlate with ASFs across a range of Si/Ge systems. Figure~\ref{fig:progression} summarizes representative descriptors and ASFs for relaxed and strained bulk Si and Ge, as well as for Si/Ge superlattices with decreasing layer thickness. We generate the strained bulk reference models by fixing the in-plane lattice parameters to those of a Si$_{0.7}$Ge$_{0.3}$ alloy, mimicking epitaxial growth on a substrate. This constraint induces tensile strain in Si and compressive strain in Ge (see Fig.~\ref{fig:progression}(b) and Supplementary Table~1). In bulk Si and Ge (Fig.~\ref{fig:progression}(a,b)), all local order parameters $Q_i^{order}$ equal unity, reflecting uniform atomic environments. For relaxed bulk systems, the bond-length descriptors satisfy $b_x \approx b_z$, due to cubic symmetry. Under strain, this symmetry is broken: $b_x$ decreases and $b_z$ increases in Si, with the opposite trend observed in Ge. These descriptor changes directly correlate with strain-induced band splittings in the corresponding SFs, consistent with the observations made in prior first-principles studies~\cite{yu2008first,hinsche2011effect,hinsche2012thermoelectric}.

Extending this analysis to Si/Ge superlattices with decreasing layer thickness (Fig.~\ref{fig:progression}(c–f)) reveals a systematic evolution of both descriptors and ASFs. Lattice mismatch in the superlattices generates internal strain, as reflected by the bond-length descriptors, which indicate that inner Si atoms experience tensile strain induced by the surrounding Ge layers. In thick-period superlattices (Si$_{26}$Ge$_{26}$ and Si$_{12}$Ge$_{12}$), inner Si atoms retain near-bulk order parameters, whereas thinner superlattices (Si$_6$Ge$_6$ and Si$_4$Ge$_4$) exhibit pronounced reductions in $Q_i$, signaling increasing deviation from bulk-like environments. Consistently, ASFs show that inner atoms in thick-period superlattices retain largely bulk-like band features (Fig.~\ref{fig:progression}(d)). 

A clear signature of the persistent internal strain in all superlattices is the splitting of the valence band maximum near the $\Gamma$ point~\cite{satpathy1988electronic,hybertsen1987theory,tserbak1993unified,proshchenko2019optimization, settipalli2020theoretical}. As the superlattice period decreases, these splittings intensify and the split states interact, leading to band mixing, hybridization, and avoided crossings along the high-symmetry paths~\cite{satpathy1988electronic,hybertsen1987theory,tserbak1993unified,proshchenko2021role}. The $\Gamma$-character of the valence band evolves accordingly: while Si$_{12}$Ge$_{12}$ retains partial Si/Ge character, Si$_{6}$Ge$_{6}$ and Si$_{4}$Ge$_{4}$ exhibit substantially stronger mixing, indicating strong dependence on heterostructure composition and layer thickness. These trends are consistent with prior work showing that Si$_6$Ge$_6$ is nearly direct~\cite{froyen1988structural}, and that related compositions, such as Si$_6$Ge$_4$, achieves a direct-gap behavior~\cite{d2012genetic}. Such progression of ASFs across these systems indicates that tuning layer thickness and composition provides a route to modifying band-gap character in Si–Ge heterostructures~\cite{froyen1988structural,d2012genetic}. While continuous band mixing is well known in random alloys~\cite{eales2019ge1}, our results demonstrate analogous behavior in layered Si–Ge heterostructures. Although stacking indirect-gap materials can enable direct-gap behavior, identifying such configurations through trial-and-error approaches is costly. The present framework offers a systematic route for exploring these design spaces, although a comprehensive classification of gap character lies beyond the scope of this work.

Interface atoms exhibit substantially reduced order parameters compared with inner atoms (Fig.~\ref{fig:progression}(e,f)), reflecting strongly perturbed local environments. Notably, interface descriptors converge across different superlattices, and their ASFs closely resemble those of inner atoms in short-period superlattices. This convergence indicates that the distinction between inner and interface environments diminishes as layer thickness approaches a few (4-6) monolayers. Together with parallel trends for Ge atoms (Supplementary Fig.~2), these results show that ASF spectral features---including strain-induced splittings, band mixing, avoided crossings, and symmetry breaking---provide a direct, atom-resolved link between local atomic environments and electronic band dispersion in semiconductor heterostructures. Note that this atom-resolved view contrasts with the total spectral functions of Si/Ge superlattices, which largely appear as superpositions of bulk-like Si and Ge bands (Supplementary Fig.~2)~\cite{satpathy1988electronic,settipalli2020theoretical}. Such total spectral functions obscure contributions from distinct local atomic environments, underscoring the advantage of ASFs as a representation for the electronic structure of heterostructures with highly heterogeneous local environments.

\subsubsection{Validation of forward model predictions} 

\begin{figure*}
\begin{center}
\includegraphics[width=\linewidth]{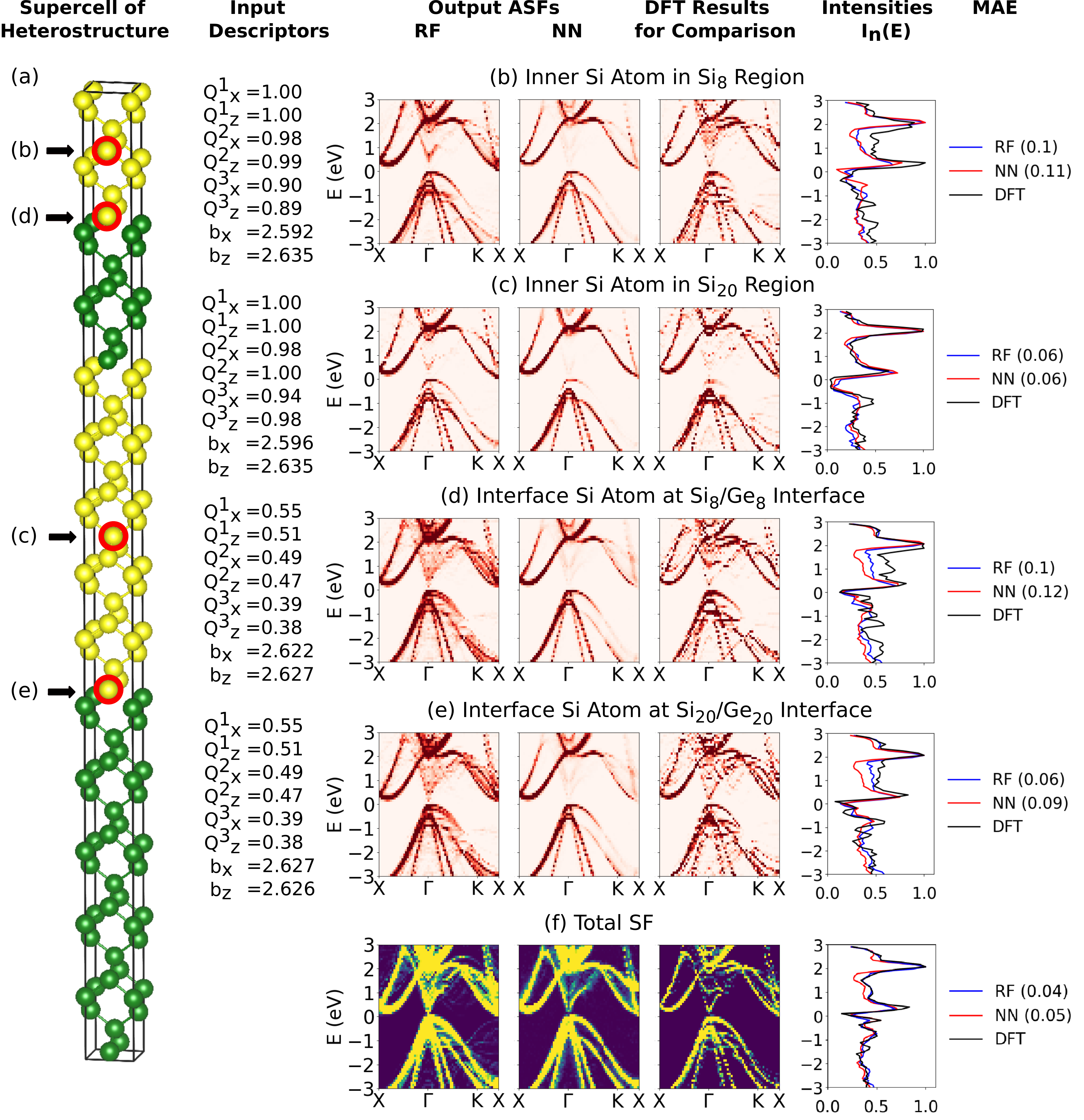}
\caption{{\bf Forward learning model predictions and validation.} (Column 1) (a) Representative supercell configuration of a test heterostructure Si$_{8}$Ge$_{8}$Si$_{20}$Ge$_{20}$, with selected Si atoms from (b) the inner Si$_{8}$ layer, (c) the inner Si$_{20}$ layer, (d) the Si$_{8}$Ge$_{8}$ interface, and (e) the Si$_{20}$Ge$_{20}$ interface regions highlighted. (Column 2) Atomic-environment descriptors used as inputs to the forward models. (Column 3-4) ASFs predicted by the RF and NN models, respectively. (Column 5) DFT-computed ASFs for comparison. (Column 6) Normalized energy-resolved intensity profiles, $I_n(E)$. (Column 7) Mean absolute errors (MAEs) for RF (blue) and NN (red) predictions. (f) Total SFs obtained by summing ASFs over all atoms in the heterostructure.
}
\label{fig:forward-test1}
\end{center}
\end{figure*}

We validate the forward learning approach using two structurally distinct systems that lie outside the training distribution: a heterogeneous multilayer structure, Si$_8$Ge$_8$Si$_{20}$Ge$_{20}$ (Fig.~\ref{fig:forward-test1}(a)), and a large-period superlattice, Si$_{28}$Ge$_{28}$ (Supplementary Fig.~3). We assess two complementary forward models---a random forest (RF) regressor and a neural network (NN)---trained using the same atomic-environment descriptors and ASF representations. The multilayer heterostructure contains Si and Ge layers of varying thicknesses and thus represents a realistic fabricated configuration. For each test system, we extract atomic-environment descriptors from DFT-optimized structures, and use as inputs to the trained models to predict the corresponding ASFs. Figure~\ref{fig:forward-test1}(b–e) compares the ASFs predicted by the RF and NN models with DFT-computed ASFs for representative Si atoms in different regions of the heterostructure. To quantify agreement, we evaluate model energy-resolved intensity profiles $I(E)$, obtained by integrating ASF intensity over momentum, instead of directly comparing two-dimensional ASF images. We obtain the normalized intensities by dividing $I(E)$ by the maximum intensity: $I_n (E) = I(E)/\text{Max}[I(E)]$ and compute the mean absolute errors (MAEs) between predicted and DFT-derived spectra: $ MAE(I_n,\hat{I}_n)=\sum_{E}|I_n(E)-\hat{I}_n(E)|/64$.

The final column of Fig.~\ref{fig:forward-test1} summarizes the normalized intensity profiles and associated errors. The computed bond-length descriptors indicate that inner Si atoms in both the thin (Si$_8$) and thick (Si$_{20}$) layers experience strained local environments. Correspondingly, their ASFs exhibit valence band splittings consistent with trends observed in superlattices (Fig.~\ref{fig:progression}). In the thinner Si$_8$ layer, reduced order parameters signal a more perturbed environment, leading to pronounced mixed Si–Ge character near the $\Gamma$ point. In contrast, the inner Si atom in the thicker Si$_{20}$ layer retains near-bulk descriptor values and exhibits an ASF closely resembling that of inner atoms in large-period superlattices. Both RF and NN models reproduce these qualitative trends. The RF model captures both mixed and bulk-like spectral features with high fidelity (MAE $\leq$ 0.11), while the NN reproduces the overall dispersion (MAE $\leq$ 0.12) but underestimates some finer Ge-like features. 

Representative Si atoms at the Si$_8$Ge$_8$ and Si$_{20}$Ge$_{20}$ interfaces exhibit similarly reduced order parameters and mixed ASFs. Both models capture the averaged spectral characteristics of these interface environments, although detailed band splittings and discontinuities are more pronounced in the DFT results. The remaining discrepancies arise primarily from the band-unfolding procedure used to compute DFT spectral functions, which becomes less well defined in heterostructures with irregular translational order. In contrast, the ML models interpolate from learned relationships between atomic environments and ASFs. We expect that incorporating additional multilayer heterostructures into the training set will further improve atom-level prediction accuracy. Despite these differences, summing the predicted ASFs over all atoms yields total spectral functions that closely match DFT results for both test systems (Fig.~\ref{fig:forward-test1}(f)). This agreement demonstrates that the forward model accurately captures the collective electronic response of complex heterostructures while retaining sensitivity to local atomic environments.

\subsection{Reverse learning model}

\subsubsection{Overview}

\begin{figure}[h!]
\begin{center}
\includegraphics[width=0.9\linewidth]{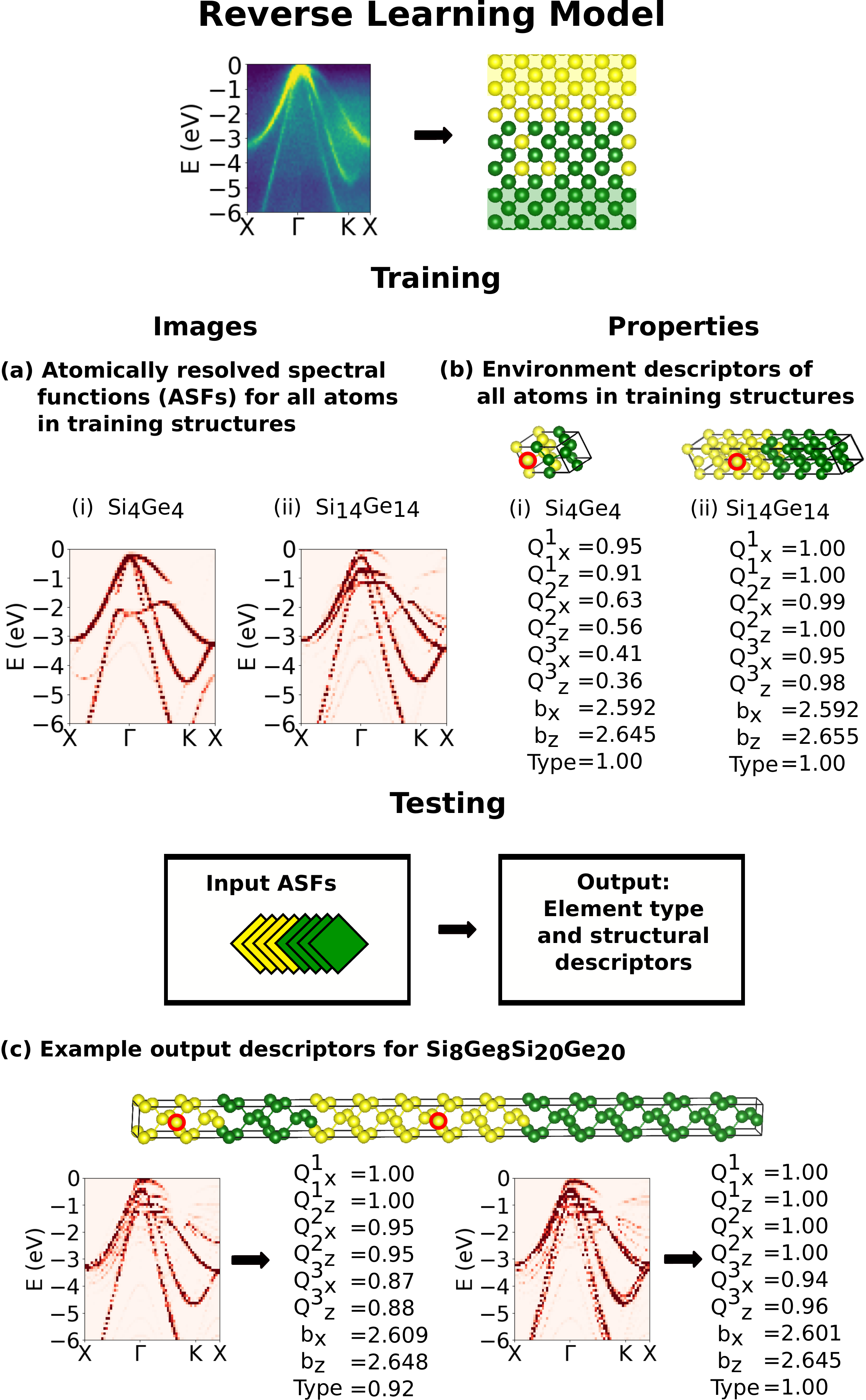}
\caption{{\bf Outline of reverse learning approach.} (a) Training images consisting of example ASFs for inner Si atoms in (i) Si$_{4}$Ge$_{4}$ and (ii) Si$_{14}$Ge$_{14}$ superlattices. (b) Atomic-environment descriptors associated with the training images, including elemental identity, effective bond lengths, and local order parameters. (c) A trained convolutional neural network (CNN) maps input ASF images to predicted atomic-environment descriptors for atoms in the heterostructure Si$_{8}$Ge$_{8}$Si$_{20}$Ge$_{20}$. Predicted descriptors are compared with those obtained directly from DFT.}
\label{fig:CNN}
\end{center}
\end{figure}

Building on the forward learning model, we develop a reverse learning model to address the inverse problem of inferring local atomic-environment descriptors directly from ASF images. The model is trained exclusively on DFT-computed ASFs but is designed to generalize to experimental angle-resolved photoemission spectroscopy (ARPES) data by explicitly incorporating variability in the training set. While direct comparisons between DFT and ARPES must be treated with care---owing to many-body effects present in experiments but absent in standard DFT---previous studies have demonstrated good agreement between DFT spectral functions and ARPES measurements in materials with weak electron–electron correlations~\cite{seo2014critical,constantinou2021fabrication,strocov2019k}. Within this context, the reverse model provides a data-driven pathway for interpreting ARPES images and extracting local structural information directly from electronic spectra. Figure~\ref{fig:CNN} outlines the reverse learning approach, in which a convolutional neural network (CNN) maps ASF images to atomic-environment descriptors. The CNN is trained using DFT-computed ASFs and corresponding descriptors for all atoms in the Si/Ge superlattices included in the forward model training set (Table~\ref{table:MLDatatable}). To assess the model’s ability to generalize beyond the training distribution, we compare CNN-predicted descriptors with ground-truth values obtained from DFT-relaxed structures not included during training.

\subsubsection{Band--dispersion--to--structure inference in a model heterostructure} 

\begin{figure}
\begin{center}
\includegraphics[width=0.95\linewidth]{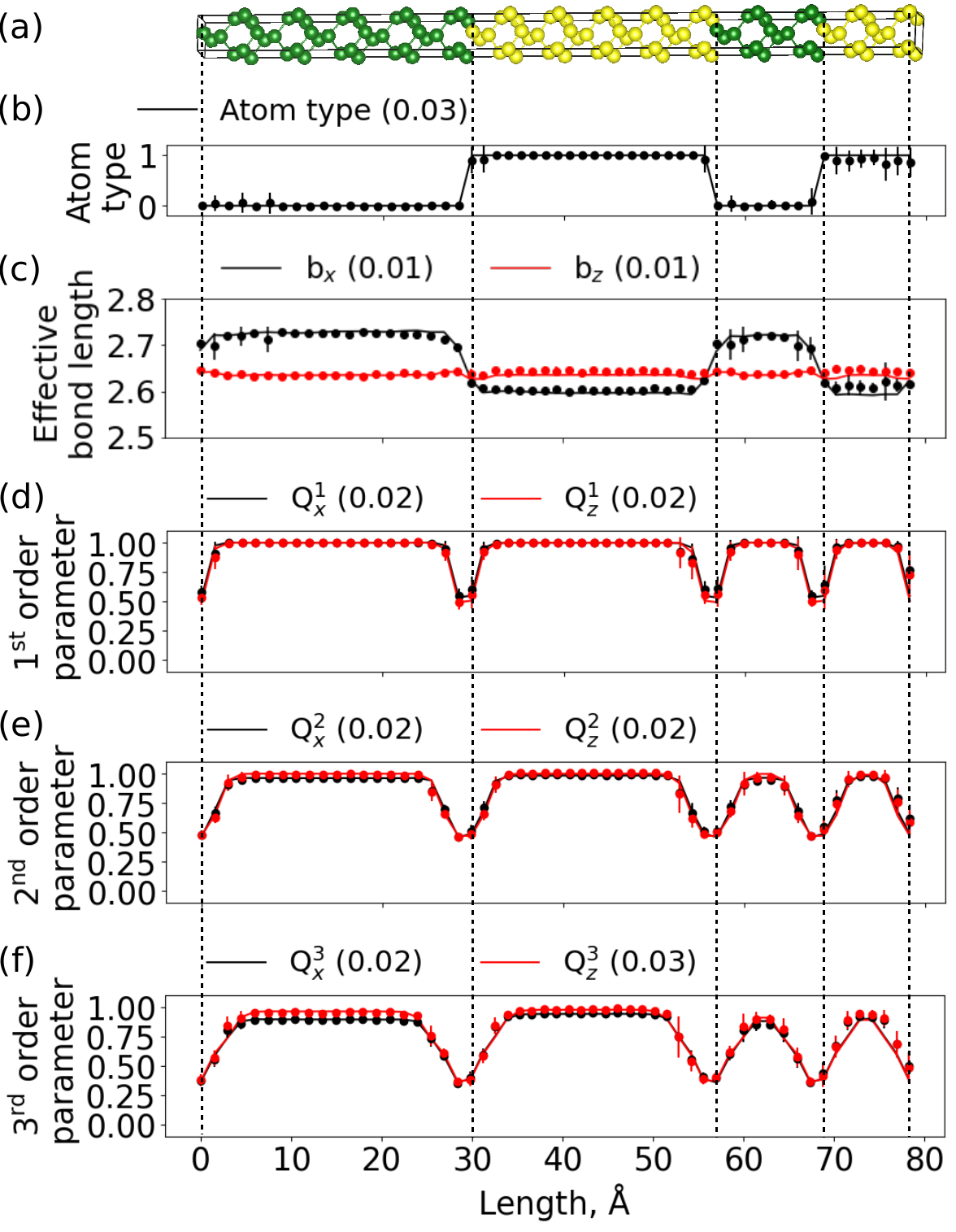}
\caption{{\bf Reverse learning model predictions for a model heterostructure.} (a) Supercell of the strain-symmetrized Si$_{8}$Ge$_{8}$Si$_{20}$Ge$_{20}$ heterostructure. Predicted (b) atomic species, (c) effective bond lengths and (d-f) spatially resolved local order parameters, $Q_i^{order}$, where $i=(x,z)$ and $order = 1,2,3$, for all atoms in the heterostructure. Predicted descriptor averages (symbols) are compared with DFT reference values (solid lines). Error bars indicate the standard deviation across input ASF images with different Fermi-level alignments. MAEs are reported in each panel.
}
\label{fig:reverse-test-heterostructure}
\end{center}
\end{figure}

We evaluate the reverse learning approach using the strain-symmetrized heterostructure Si$_{8}$Ge$_{8}$Si$_{20}$Ge$_{20}$ and superlattice Si$_{28}$Ge$_{28}$ (Supplementary Fig.~4), both of which are also used to validate the forward model. Figure~\ref{fig:reverse-test-heterostructure}(a) shows the heterostructure supercell, while Fig.~\ref{fig:reverse-test-heterostructure}(b–f) compares CNN-predicted atomic-environment descriptors (symbols) with DFT reference values (solid lines). For each atom, we generate multiple input ASF images by applying different Fermi-level alignments (Supplementary Fig.~11) and synthetic variations in brightness and noise. The predicted descriptors are summarized by their mean values, with error bars indicating the standard deviation across the ensemble. Prediction accuracy for each descriptor $D$ is quantified using the mean absolute error, $ MAE(D, \hat{D})=\frac{1}{(p\times n)}\sum_{i}^{p\times n}|D_i-\hat{D}_i|$, where $p$ is the number of atoms and $n$ is the number of applied Fermi level shifts ($n=13$). The reverse model captures spatial variations of atomic-environment descriptors across the heterostructure with high fidelity. Atomic species types are predicted most accurately for bulk-like inner atoms in thicker layers, with reduced accuracy near interfaces and in thinner layers. Predicted effective bond lengths, $b_x$ and $b_z$ (Fig.~\ref{fig:reverse-test-heterostructure}(c)), reflect the distinct strain environments of Si and Ge layers. Because all layers share the same in-plane lattice constant, $b_z$ remains nearly constant ($\sim$2.64 \AA), whereas $b_x$ varies more strongly due to strain-induced changes in cross-plane spacing~\cite{proshchenko2021role}. As expected, $b_x$ is larger in Ge layers ($\sim$2.72 \AA) than in Si layers ($\sim$2.60 \AA), consistent with bulk values (Ge: 2.73 \AA\ and Si: 2.58 \AA) and residual internal strain.
 
The predicted order parameters (Fig.~\ref{fig:reverse-test-heterostructure}(d–f)) clearly delineate interface regions, where reduced same-species coordination leads to lower values. Inner regions of the thicker Si$_{20}$ and Ge$_{20}$ layers exhibit order parameters close to unity, indicating bulk-like environments. Higher-order parameters ($Q_i^2$ and $Q_i^3$) resolve variations within thinner layers more effectively than $Q_i^1$. Across all descriptors, prediction accuracy is highest in bulk-like regions and lowest near interfaces, consistent with the increased complexity of interface ASFs (Fig.~\ref{fig:forward-test1}(d-e)) and the limited representation of such environments in the training data. Note that the descriptors are derived from Voronoi tessellations and are therefore sensitive to small atomic displacements~\cite{leonardi2012strain,garg2023grain}. Despite this sensitivity, the consistently small uncertainties demonstrate that the CNN model learns robust relationships between ASF features and atomic environments, and remains resilient to noise, contrast variations, and artifacts commonly encountered in ARPES measurements, as well as to limitations of the underlying DFT calculations. Together, these results show that the reverse learning model can reliably infer atomic-scale structural information directly from electronic band dispersion images of complex semiconductor heterostructures.

\subsubsection{Band--dispersion--to--structure inference in bulk silicon}

We further evaluate the reverse learning model by testing its ability to infer atomic environments from band dispersion images of bulk Si. In bulk Si, the SF and ASFs are identical, as all atoms share the same local environment. Notably, the model is trained exclusively on ASFs of Si/Ge superlattices and SFs of pristine bulk Si or Ge are not included in the training set. We compute SFs for both relaxed and strained bulk Si supercells using DFT. The strained configuration corresponds to epitaxial growth on a Si$_{0.7}$Ge$_{0.3}$ alloy substrate, inducing a 1.73\% tensile strain along the in-plane lattice direction $a^\prime$. Supplementary Fig.~5 show the SFs of relaxed and strained Si, respectively, plotted along the $X-\Gamma-K-X$ path of the bulk Si Brillouin zone. As expected, the strained case exhibits a splitting of the valence band maxima near $\Gamma$. As in the heterostructure tests, we generate ensembles of input images with varied Fermi level alignments and synthetic noise to assess model robustness.

We next test the model using an experimental ARPES image of a Si thin film, adapted from from Fig.~6.2 of Ref.~\citenum{constantinou2021fabrication}. The ARPES spectra along different symmetry directions was originally presented as separate panels. We combine the panels into a single image and use it as input to the trained model. We also apply additional vertical shifts to mimic variations in Fermi level alignments. Supplementary Fig.~5 summarizes the predicted atomic descriptors for relaxed bulk Si, strained bulk Si, and the ARPES image. In all cases, the model correctly identifies the atom type as Si and predicts order parameters close to unity, consistent with bulk-like atomic environments without interfaces or compositional mixing. The predicted bond-length descriptors reflect the symmetry and strain state encoded in the band dispersion. For relaxed bulk Si and the ARPES image, the model predicts $b_x \approx b_z$, indicating a high-symmetry environment. For strained bulk Si, it predicts $b_z \textgreater b_x$, consistent with tensile strain along $a^{\prime}$, imposed by the epitaxial substrate. Although the absolute values of $b_x$ and $b_z$ are systematically slightly higher than those obtained directly from DFT, the model accurately reproduces all strain-induced trends. Tests on additional strained bulk Si configurations (Supplementary Fig.~6 and Supplementary Table~4) further confirm this behavior. Together, these results demonstrate that the reverse learning model can infer atomic-scale structural information directly from band dispersion images, including experimental ARPES data that lie outside the training distribution.

\begin{figure}
\begin{center}
\includegraphics[width=0.87\linewidth]{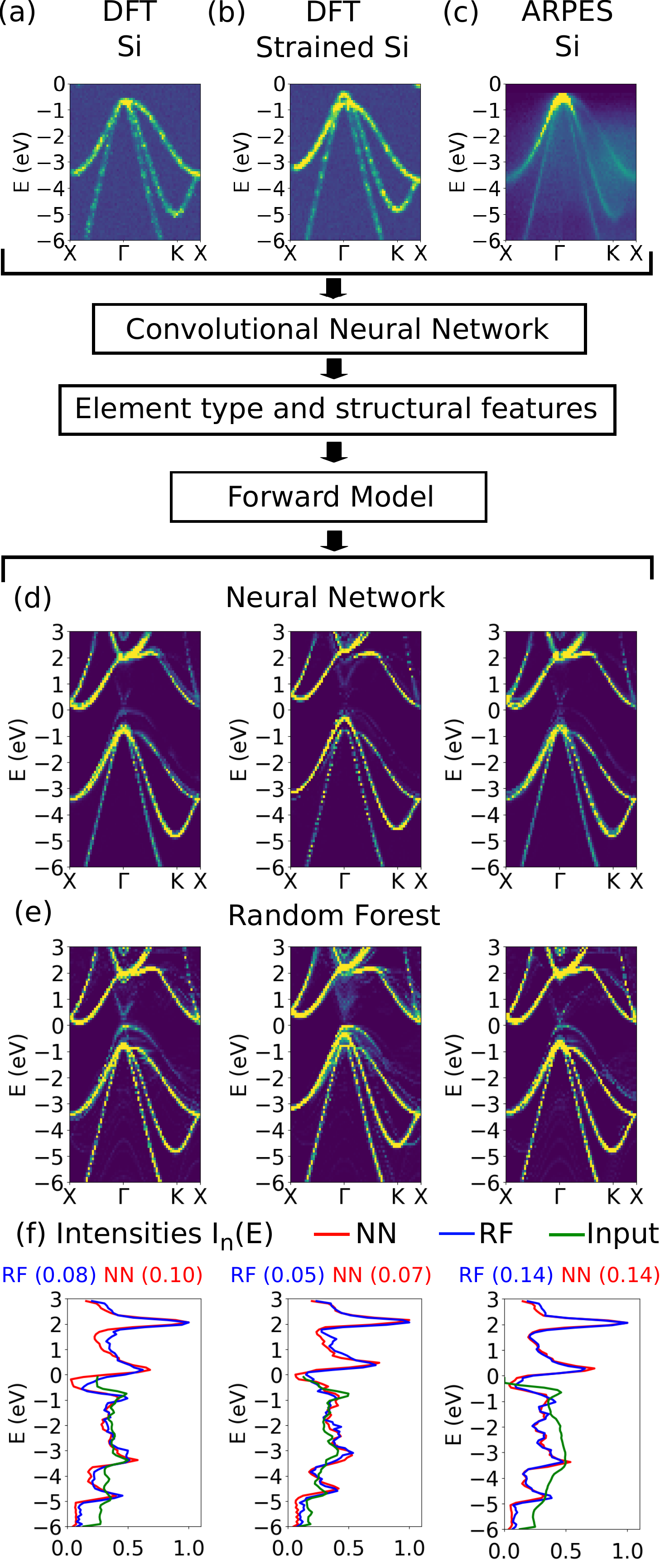}
\caption{{\bf Coupled forward–reverse learning framework and self-consistent validation.} Input spectral images include DFT SFs for (a) relaxed and (b) strained bulk Si, and (c) ARPES spectra of a Si thin film~\cite{constantinou2021fabrication}. The trained CNN model extracts environment-environment descriptors from the input images, which are then passed to the forward learning model. Panels (d) and (e) show spectral functions reconstructed by the neural network (NN) and random forest (RF) forward models, respectively, using the inferred descriptors. (f) Comparison of normalized energy-resolved intensities between reconstructed and input SFs, respectively, with MAEs indicated.
}
\label{fig:combined-model}
\end{center}
\end{figure}

\subsection{Closed-loop band--dispersion--to--atomic--structure inference}

Finally, we implement a bidirectional band--dispersion--to--structure inference framework by coupling the forward and reverse learning models. This coupling establishes a closed-loop workflow in which atomic environments are inferred from spectral images and then used to reconstruct the corresponding SFs, enabling self-consistent validation against the input spectra. Figure~\ref{fig:combined-model} illustrates this workflow using the bulk Si systems discussed in the previous subsection. In this framework, the CNN first maps the input SF images to atomic-environment descriptors, which are subsequently passed to the trained forward learning models based on random forests (RF) and neural networks (NN). Figures~\ref{fig:combined-model}(d,e) show the SFs reconstructed by the forward models. Because the forward model is trained on DFT-computed ASFs that include both valence and conduction bands, it predicts the full band structure, including conduction band features that may be absent in the original ARPES images. The forward models, particularly the RF model, also reproduce faint band-mixing features near the $\Gamma$ point, reflecting patterns learned from the superlattice ASF training data. Figure~\ref{fig:combined-model}(f) shows close agreement between reconstructed and input spectra for both DFT-computed and experimental ARPES bulk Si images. Larger errors observed for the ARPES case arise primarily from differences in image quality, contrast, and resolution relative to the training data. Despite these differences, the closed-loop framework consistently links electronic band dispersion and atomic structure in a self-consistent manner. This capability is particularly relevant for interpreting ARPES measurements in systems such as $\delta$-doped (As- or P-doped) semiconductors, where conduction-band states may be absent prior to doping but emerge upon dopant incorporation~\cite{constantinou2021fabrication}. In such cases, the coupled framework offers a data-driven route for inferring hidden or weak spectral features and relating them to their underlying atomic-scale structural origins.

\section*{Discussion}

In this work, we show that electronic spectral signatures are intrinsically linked to atomic-scale environments and introduce an ML-assisted framework that identifies and exploits the patterns encoding this connection. We represent electronic structure using ASFs, which capture how specific local atomic environments contribute to band features, including strain-induced splittings, band mixing, avoided crossings, and changes in Bloch character. By explicitly resolving contributions from inner and interfacial atoms, the ASF representation goes beyond global band descriptions and preserves local structural context within a unified momentum-space framework. The local, information-dense description enables the models to accurately learn and infer from a limited number of training structures while preserving interpretability. Using this representation, the forward learning model establishes how atomic-environment descriptors map to electronic band dispersion. The reverse learning model further demonstrates that atomic-scale structural information can be inferred directly from band dispersion images, including experimental ARPES data, despite being trained exclusively on DFT-computed spectral functions. Spectral fingerprints alone allow the model to distinguish bulk-like environments from interfacial or strongly perturbed configurations. This capability moves beyond prior ML methods that focus primarily on band reconstruction or latent representation learning, enabling instead a direct, physics-informed interpretation of electronic spectra. When coupled with the forward model, these capabilities form a closed loop, in which inferred atomic descriptors are used to reconstruct SFs and directly compared with the input images, providing a self-consistent validation of the learned relationships. While this study focuses on Si/Ge heterostructures and selected high-symmetry paths, the results illustrate what becomes possible when electronic bands are treated as learnable and decomposable objects rather than monolithic outputs of simulation or experiment. At the same time, this work represents an initial step. Extending the framework to broader materials classes, incorporating more complex band topologies, and accounting for many-body effects beyond standard DFT remain important directions for future work.

Beyond its conceptual implications, the framework provides practical capabilities relevant to computational and experimental materials research, including direct comparison between first-principles calculations and ARPES measurements, physics-informed interpretation of electronic spectra, and layer-resolved analysis of heterostructure band contributions. Within this context, prospects such as engineering direct-gap behavior from indirect-gap constituents or identifying weak spectral features that may be inaccessible to conventional ARPES measurements emerge naturally. By demonstrating that atomic environments can be inferred from, and used to reconstruct, electronic band dispersion in a self-consistent manner, this study establishes a foundation for data-driven, physics-informed exploration and inverse design of complex electronic materials. More broadly, this work contributes to ongoing efforts to establish scalable, representation-centric approaches that may ultimately support transferable, foundation-level models of electronic structure linking atomic configuration, spectral response, and materials functionality within a unified computational framework.

\section*{Data availability}

All data generated or analysed during this study are included in this published article and its supplementary material files. Example datasets generated and/or analyzed during the current study are available in the CUANTAMLab public GitHub repository~\cite{githubcuantam} [\href{https://github.com/CUANTAM/Spectral-Function-Dataset}{url}].



\section*{Acknowledgements}
We gratefully acknowledge funding from the Defense Advanced Research Projects Agency (Defense Sciences Office) [Agreement No.: HR0011-16-2-0043]. We acknowledge funding from the National Science Foundation Harnessing the Data Revolution NSF-HDR-OAC-1940231. This work utilized the Summit supercomputer, which is supported by the National Science Foundation (awards ACI-1532235 and ACI-1532236), the University of Colorado Boulder, and Colorado State University. The Summit supercomputer is a joint effort of the University of Colorado Boulder and Colorado State University.

\section*{Author contributions}

\noindent A.K.P contributed to the acquisition and the analysis of data and the creation of new scripts used in the study. S.N. contributed to the conception and the design of the work, the interpretation of data, drafting and revision of the article.

\section*{Competing interests}

\noindent The authors declare no competing interests.

\section{Methods}

\subsection*{Training and test structures for all ML models}

We design the training dataset to span a physically relevant range of Si/Ge heterostructures. The dataset consists of ideal Si$_n$Ge$_n$ superlattices with atomically sharp interfaces, where $n$ denotes the number of Si and Ge monolayers stacked along the [001] direction.

\subsubsection*{Training Superlattices} 

We include superlattices with both even and odd numbers of monolayers to sample layer periodicity effects. Even-period superlattices are denoted as Si$_{2p}$Ge$_{2p}$ with $(p=1,2,\dots, 14)$,  while odd-period superlattices are denoted as Si$_{2q-1}$Ge$_{2q-1}$Si$_{2q-1}$Ge$_{2q-1}$ with $(q=1,2,\dots, 7)$. The corresponding supercells (SCs) contain $4p$ and $4(2q-1)$ atoms, respectively. All SCs are generated from a four-atom tetragonal template; for odd-period superlattices, we double the SC size along [001] to correctly preserve periodicity. To probe strain effects, we include both strain-symmetrized and strained superlattices with in-plane strains of 0.00\%, 0.59\%, 1.16\%, 1.73\%, and 2.31\%, defined relative to the bulk Si lattice constant: $((a^\prime-a_{Si})/a_{Si})\times 100$, where $a_\text{Si}$ = 5.47 \AA. These strain values correspond to epitaxial growth on Si$_{1-x}$Ge$_x$ alloy substrates with Ge concentrations: $x = 0, 0.1, 0.2, 0.3,$ and $0.4$. 

\subsubsection{Forward and reverse model test structures} 

We evaluate the forward model on strain-symmetrized structures excluded from training to assess generalization: a Si$_{8}$Ge$_{8}$Si$_{20}$Ge$_{20}$ heterostructure and a Si$_{28}$Ge$_{28}$ superlattice. The reverse model is tested on the same structures, along with relaxed and strained bulk Si models and experimental ARPES images from Ref.~\cite{constantinou2021fabrication}.

\begin{figure}
\begin{center}
\includegraphics[width=1.0\linewidth]{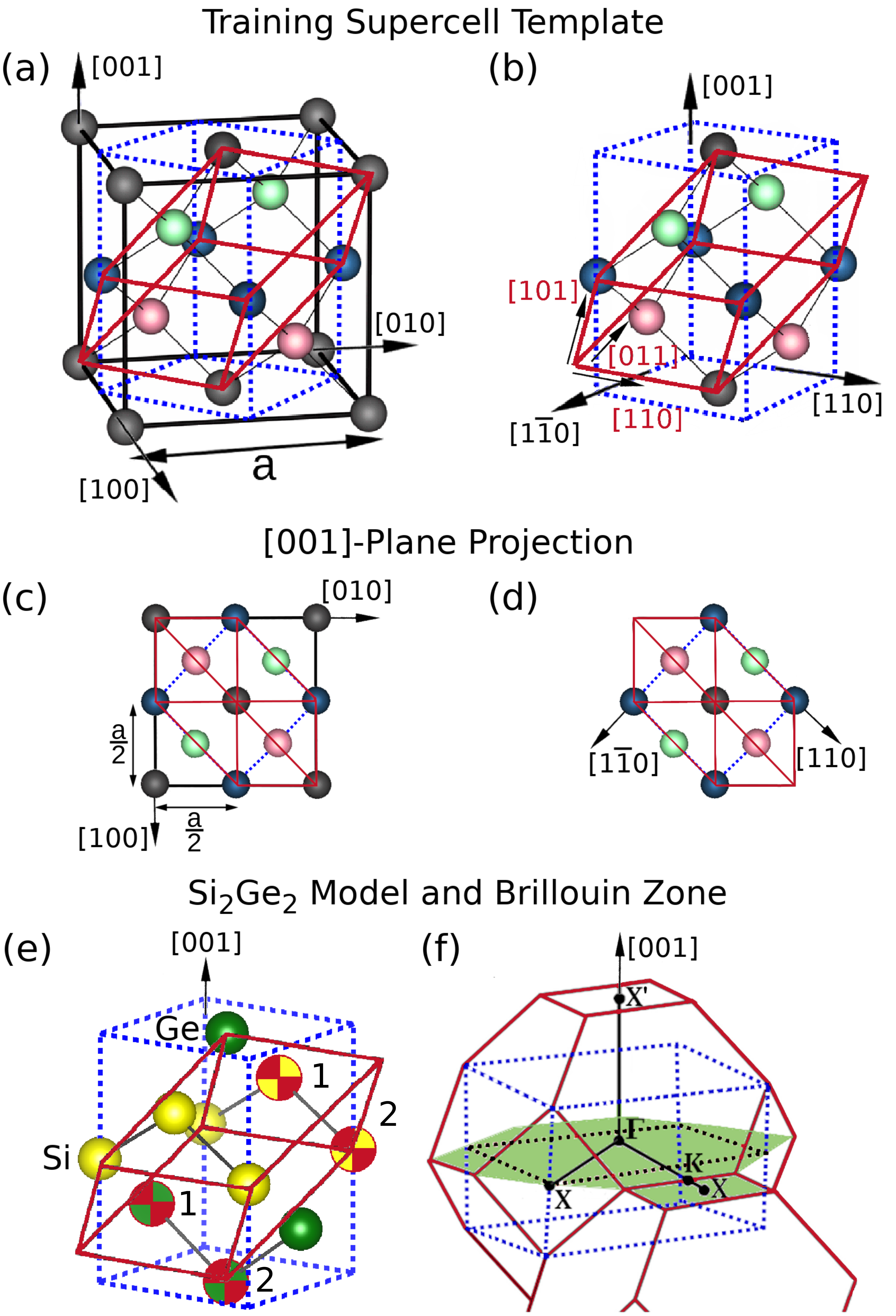}
\caption{{\bf Generation of training supercells (SCs) and selection of reference cells (RCs):} (a,b) Tetragonal SC template (blue dashed lines) derived from a bulk Si conventional cell (black solid lines), where $a$ denoted the bulk Si lattice constant. The SC template contains four atomic positions, corresponding to one position per monolayer stacked along the [001] direction, indicated in black, red, blue, and green. A representative two-atom RC selected for band unfolding is shown by solid red lines. (c,d) Atomic positions in Si conventional cell, SC template and RC projected along the [001] direction. (e) Si$_2$Ge$_2$ superlattice SC, with two Si and two Ge atoms highlighted in red, while all other atoms represent periodic replicas. Pairs of marked SC atoms (`1' and `2' ) are mapped to the two corresponding RC atomic positions, respectively. (f) Brillouin zones (BZs) of the SC (blue) and RC (red). The black dashed line indicates the projection of the SC BZ onto the [001] plane passing through the $\Gamma$ point. Symmetry points and paths in the green plane are used to obtain effective band structures or spectral functions.}
\label{fig:SLcell}
\end{center}
\end{figure}

\subsubsection*{Supercell construction} 

All SCs are generated from a four-atom tetragonal Si template (Si$_4$) derived from the conventional cubic cell (Fig.~\ref{fig:SLcell}). The optimized lattice parameters are $a^{\prime}=b^{\prime}=2.73$\AA\ and $c=5.47$\AA, in agreement with previous DFT results~\cite{wright2006density}, noting that DFT typically overestimates experimental Si lattice constants by approximately 1\%~\cite{semiconductor2012general}. The template has half the volume of the cubic cell, with optimized lattice parameters $a^{\prime}=b^{\prime}=2.73$\AA\ and $c=5.47$\AA. The template can be used to span Si systems with cubic symmetry, e.g., [001] grown superlattices, by replicating in the [$110$], [$1\bar{1}0$] and [$001$] directions. This template allows us to investigate a large variety of superlattices and heterostructures, while keeping the computational expense at a minimum. The template contains four atomic planes separated by $a/4$ along [001]; superlattices are constructed by assigning Si and Ge atoms to these planes and replicating the template as needed. Geometry optimization yields strain-symmetrized or strained configurations. Lattice parameters for all optimized SCs are reported in Supplementary Table 1.


\subsubsection*{Combined model test structures and ARPES images} 

The combined forward–reverse model is tested on bulk Si systems modeled using the Si$_4$ supercell, including a strained configuration corresponding to growth on a Si$_{0.7}$Ge$_{0.3}$ substrate (1.73\% strain). Experimental ARPES spectra are adopted from Fig.~6.2 of Ref.~\citenum{constantinou2021fabrication}. We combine the band dispersions along the $\Gamma-X$ (Fig.~6.2(a)) and $\Gamma-K-X$ paths (Fig.~6.2(c)) into a single image. We interpolate the experimental images from the resolution provided in Ref.~\cite{constantinou2021fabrication} to 64 $\times$ 64 pixels over an energy window from -6 eV to 0 eV. 

\subsection*{DFT computation details}

We optimize lattice constants and atomic positions of all training and test supercells using the conjugate gradient algorithm~\cite{press1996numerical}. We sample the SC BZ with an 
$11\times11\times11$ Monkhorst–Pack $k$-point mesh~\cite{monkhorst1976special}, which ensures adequate sampling along the [001] direction for heterostructures with uneven Si and Ge layer thicknesses. To simulate applied strain, we fix the in-plane lattice constants ($a^\prime$ and $b^\prime$) to the substrate values and relax the cell along the [001] direction. All DFT calculations are performed using the OpenMX code~\cite{ozaki2003variationally,ozaki2004numerical,ozaki2005efficient,openmx2013}, which employs norm-conserving pseudopotentials generated with multiple reference energies~\cite{morrison1993nonlocal} and linear combination of optimized pseudoatomic basis functions~\cite{ozaki2003variationally}. We use the Perdew–Burke–Ernzerhof exchange–correlation functional~\cite{perdew1996generalized} within the generalized gradient approximation. Self-consistent field (SCF) calculations are performed during the geometry optimization with energy convergence threshold set to $10^{-9}$ Hartree. The SCs are optimized until the maximum force on an atom became less than $10^{-4}$ Hartree Bohr$^{-1}$. A regular mesh of 200 Ryd in real space is used for the numerical integrations and solution of Poisson equation~\cite{soler2002siesta}.

We neglect spin–orbit coupling, as strain-induced band splittings in Si/Ge heterostructures exceed spin–orbit splittings~\cite{hybertsen1987theory}. For the Si and Ge atoms, 2, 2, and 1 optimized radial functions are allocated for the s-, p- and d-orbitals, respectively, as denoted by s2p2d1. The one-particle wave functions are expressed by the linear combination of pseudo-atomic orbital (PAO) basis functions centered on atomic site ~\cite{ozaki2003variationally,ozaki2004numerical}. A cutoff radius of 7.0 Bohr was used for all the basis functions. Following relaxation, we perform non self-consistent field (NSCF) calculations using the linear combinations of atomic orbitals (LCAO) pseudopotential method~\cite{ozaki2003variationally,ozaki2004numerical}. We obtain the eigenstates and energy for the range from -10 eV to 10 eV. We use a $7\times 7\times 7$ k-point mesh generated according to the Monkhorst-Pack method~\cite{monkhorst1976special} to sample the supercell BZ. Such k-point mesh has been used in DFT studies for calculation of electronic structure of two-atom Si lattice~\cite{bystrom2019pawpyseed}.

\subsection{Calculation of Atomic Environment Descriptors} 

The predictive performance of ML models for materials properties critically depends on the choice of descriptors~\cite{ward2017including,ghiringhelli2015big}. Prior work showed that local structural descriptors dominate predictive accuracy in ML models of electronic transport in Si/Ge superlattices, reflecting the strong sensitivity of transport properties to atomic-scale environments~\cite{proshchenko2019optimization,proshchenko2021role,settipalli2020theoretical,schaffler1997high,proshchenko2019heat}. In contrast, elemental-property descriptors contribute minimally in binary systems~\cite{ward2017including}. Guided by these findings, we adopt physics-informed random forest (RF) and neural network (NN) models that combine one elemental and several structural descriptors~\cite{pimachev2021first}. Each atom $X$ is represented using one elemental descriptor (Si = 1, Ge = 0) and two classes of structural descriptors derived from Voronoi tessellations and crystal graphs~\cite{pimachev2021first}: direction-dependent effective bond lengths and local structural order parameters. Together, these descriptors capture bonding anisotropy and deviations from local structural order across Si/Ge superlattices. The resulting descriptor set distinguishes atomic environments in both bulk-like and interfacial regions. Representative descriptor values for selected superlattices are reported in Supplementary Tables~2 and~3. Full mathematical definitions and implementation details are provided in the Supplementary Information.

The crystal-graph-based descriptors used here are conceptually related to graph neural network representations~\cite{xie2018crystal,choudhary2021atomistic,gupta2024structure}, but differ in design philosophy. Rather than learning node features from large datasets, we define physically motivated node descriptors a priori, enabling accurate learning from limited data while preserving interpretability, which is critical given the scarcity of electronic transport data of heterostructures in existing DFT databases~\cite{jain2013commentary,choudhary2020joint}.

\subsection*{Supercells and reference cells}

We compute SFs by unfolding SC electronic band structures into the extended-zone representation of a common reference cell (RC), enabling direct comparison across superlattices and heterostructures with different compositions and periods~\cite{popescu2012extracting,boykin2007approximate,boykin2007brillouin,boykin2009non,lee2013unfolding,lee2020unfolding}. Although RC identification becomes nontrivial in systems containing interfaces, RCs can be mathematically defined and used for unfolding regardless of structural complexity~\cite{chen2018layer}. For consistency, we select a two-atom, primitive-like RC resembling the FCC primitive cell of Si; however, the unfolding framework remains general and valid irrespective of RC choice. Despite variations in SC lattice vectors and Brillouin zones, all RCs contain two lattice sites, providing a uniform basis for unfolding. Figure~\ref{fig:SLcell}(a–d) illustrates a representative rhombohedron RC (red solid lines) embedded within the SC template (blue dashed lines) and the conventional cubic cell (black solid lines). The RC has one quarter of the conventional-cell volume and reduces to the primitive cell of FCC Si in the absence of symmetry breaking. For each superlattice or heterostructure, we construct the RC directly from the SC lattice vectors via a linear transformation that preserves translational symmetry along the [001] growth direction. We use the resulting RC BZ to compute ASFs along selected high-symmetry paths, which serve as training data for the ML models. The RCs do not necessarily correspond to irreducible primitive cells, but provide a consistent mathematical basis for unfolding across all systems studied. Full details of the SC–RC transformation and basis construction are provided in the Supplementary Information.

\subsection{Spectral weights and spectral functions}

Electronic band structures of superlattices and heterostructures, while readily accessible via DFT, are difficult to interpret due to structural diversity and band folding inherent to supercell (SC) models. Similar challenges arise in random alloys, defect systems, and heterostructures~\cite{popescu2010effective,boykin2007approximate,popescu2012extracting,boykin2007brillouin,boykin2009non,chen2018layer}. As a result, raw SC band structures vary strongly with size and composition (Supplementary Fig.~1), limiting their usefulness as training data for ML models. To address this challenge, we adopt the effective band structure or spectral function (SF) formalism\cite{popescu2010effective,boykin2007approximate,popescu2012extracting,boykin2007brillouin,boykin2009non,chen2018layer,ku2010unfolding,lee2013unfolding}, which unfolds SC band structures into an extended-zone representation of a common reference cell (RC). This approach enables direct comparison across systems with different periods and compositions, and maintains direct correspondence with ARPES measurements, making SFs well suited for training both forward and reverse ML models.

We obtain SC eigenstates from non-self-consistent DFT calculations and unfold onto RC Bloch states along the $X-\Gamma-K-X$ path of the RC BZ. The key methodological advance in this work is the construction of the ASFs (Supplementary Eq.~15). ASFs decompose the unfolded SF into contributions from individual atoms, enabling us to directly link local atomic environments to features in the electronic band structure. This atomic resolution is essential for training forward and reverse learning models that map between structure and electronic response. We evaluate ASFs over an energy window from -6 to 3 eV using Gaussian-broadened delta functions (width 0.02 eV). The resulting  $A^p(k,E)$ maps are interpolated onto fixed-size grids and used as training data for the ML models. Total SFs are recovered by summing ASFs over all atoms. Full mathematical definitions of spectral weights, unfolding expressions, and the construction of orbitally and atomically resolved SFs are provided in the Supplementary Information.

\subsection{ML Model Implementations}

We implement three machine-learning models: (i) forward learning models that predict ASFs from atomic descriptors, (ii) a reverse learning model that infers descriptors from ASF images, and (iii) a combined forward–reverse model that links structure and electronic response. We implement the forward learning task using both a neural network (NN) (Supplementary Table~5) and a random forest (RF) regressor (Supplementary Fig.~7). Both models take nine atomic descriptors as input and predict interpolated ASF values $A^p(k,E)$ on a fixed grid. The NN maps descriptors directly to ASF images, while the RF provides a complementary, interpretable ensemble-based baseline. The NN is trained using the MAE loss with the Adam optimizer (Supplementary Fig.~8), and the trained models are used to predict ASFs for previously unseen structures. We implement the reverse learning task using a convolutional neural network (CNN) (Supplementary Table~6) that takes ASF images as input and predicts the corresponding atomic descriptors. The CNN extracts spatial patterns in ASF images associated with symmetry breaking and interfacial effects in heterostructures. To improve robustness against experimental variability, we train the CNN on ASFs with multiple Fermi-level alignments (see the subsection ``Effects of Fermi-level alignment" in Supplementary Material and Supplementary Fig.~10-11). The reverse model is optimized using MAE loss and the Adam optimizer (Supplementary Fig.~9). Finally, we combine the trained forward and reverse models into a unified framework that enables bidirectional mapping between atomic structure and electronic response. Full model architectures, hyperparameters, training schedules, and implementation details are provided in the Supplementary Information.

\bibliographystyle{naturemag}
\bibliography{MLLiterature}

\end{document}